\documentclass[%
 reprint,
superscriptaddress,
 amsmath,amssymb,
 aps,
]{revtex4-1}

\usepackage{amssymb}
\usepackage{graphicx}
\usepackage{dcolumn}
\usepackage{bm}
\usepackage[colorlinks=true, allcolors=blue]{hyperref}
\usepackage{color}
\usepackage{verbatim}
\newcommand{\q}{\mathbf{q}}
\newcommand{\p}{\mathbf{p}}
\newcommand{\K}{\mathcal{K}}

\begin{document}

\preprint{APS/123-QED}

\title{Interplay between in-plane and flexural phonons in electronic \\ transport of two-dimensional semiconductors}

\author{A. N. Rudenko}
\email{a.rudenko@science.ru.nl}
\affiliation{\mbox{Key Laboratory of Artificial Micro- and Nano-structures of Ministry of Education and School of Physics and Technology,} Wuhan University, Wuhan 430072, China}
\affiliation{\mbox{Institute for Molecules and Materials, Radboud University, Heijendaalseweg 135, NL-6525 AJ Nijmegen, Netherlands}}
\affiliation{\mbox{Theoretical Physics and Applied Mathematics Department,
Ural Federal University, 620002 Ekaterinburg, Russia}}
\author{A. V. Lugovskoi}
\affiliation{\mbox{Institute for Molecules and Materials, Radboud University, Heijendaalseweg 135, NL-6525 AJ Nijmegen, Netherlands}}
\author{A. Mauri}
\affiliation{\mbox{Institute for Molecules and Materials, Radboud University, Heijendaalseweg 135, NL-6525 AJ Nijmegen, Netherlands}}
\author{Guodong Yu}
\affiliation{\mbox{Key Laboratory of Artificial Micro- and Nano-structures of Ministry of Education and School of Physics and Technology,} Wuhan University, Wuhan 430072, China}
\affiliation{\mbox{Institute for Molecules and Materials, Radboud University, Heijendaalseweg 135, NL-6525 AJ Nijmegen, Netherlands}}
\author{Shengjun Yuan}
\email{s.yuan@whu.edu.cn}
\affiliation{\mbox{Key Laboratory of Artificial Micro- and Nano-structures of Ministry of Education and School of Physics and Technology,} Wuhan University, Wuhan 430072, China}
\affiliation{\mbox{Institute for Molecules and Materials, Radboud University, Heijendaalseweg 135, NL-6525 AJ Nijmegen, Netherlands}}
\author{M. I. Katsnelson}
\affiliation{\mbox{Institute for Molecules and Materials, Radboud University, Heijendaalseweg 135, NL-6525 AJ Nijmegen, Netherlands}}
\affiliation{\mbox{Theoretical Physics and Applied Mathematics Department,
Ural Federal University, 620002 Ekaterinburg, Russia}}

\date{\today}

\begin{abstract}
Out-of-plane vibrations are considered as the dominant factor limiting the intrinsic carrier mobility of suspended two-dimensional materials at low carrier concentrations. Anharmonic coupling between in-plane and flexural phonon modes is usually excluded from the consideration. Here we present a theory for the electron-phonon scattering, in which the anharmonic coupling between acoustic phonons is systematically taken into account. Our theory is applied to the typical group V two-dimensional semiconductors: hexagonal phosphorus, arsenic, and antimony. We find that the role of the flexural modes is essentially suppressed by their coupling with in-plane modes. At dopings lower than 10$^{12}$ cm$^{-2}$ the mobility reduction does not exceed 30\%, being almost independent of the concentration.
{Our findings suggest that compared to in-plane phonons, flexural phonons are considerably less important in the electronic transport of two-dimensional semiconductors, even at low carrier concentrations.}

\end{abstract}

\maketitle

\section{\label{sec1}Introduction}
Charge carrier transport in two-dimensional (2D) materials is different from their three-dimensional analogs \cite{Katsnelson-Book, Avouris}. 
Among other reasons \cite{Peres,DasSarma}, this is mainly related to the existence of the flexural phonon modes (out-of-plane vibrations), whose energy dispersion is quadratic in wave vector compared to the linear dispersion of conventional (in-plane) modes \cite{Katsnelson-Book,Lifshitz,AccChemRes}. As the charge carrier scattering on phonons is the main factor limiting the intrinsic mobility, understanding of the interplay between flexural and in-plane phonons is of vital importance for 2D material's electronic transport.

The contributions of in-plane and flexural phonons are usually considered independently of each other, at the level of the harmonic approximation. Moreover, the role of flexural phonons is often completely neglected \cite{Kaasbjerg2013,Qiao,InSe}, assuming that their contribution is suppressed by the presence of a substrate \cite{Amorim}. At the same time, flexural phonons are considered to play the dominant role in the electron-phonon scattering of free-standing graphene \cite{Morozov2008,Mariani2008,Castro2010b}, and may contribute significantly in the presence of a dielectric substrate \cite{Gunst,Sohier17}. Transport properties of conventional 2D semiconductors like black phosphorus are less affected by flexural modes in the regime of high doping \cite{Rudenko2016}, whereas at low dopings the situation is unclear. Furthermore, flexural modes could significantly renormalize the electronic spectrum, leading to the formation of bound states \cite{Flexuron,Brener}. 

The separation between the in-plane and flexural modes becomes undefined at wave vectors $q\lesssim q^*$, smaller than the characteristic vector $q^*=\sqrt{3TY/16\pi\kappa^2}$, determined by the Young modulus $Y$, flexural rigidigy $\kappa$, and temperature $T$ \cite{AccChemRes,Nelson-orig,Nelson}. This is a manifestation of the anharmonic coupling regime, which is typical to low carrier concentrations with small Fermi wave vectors, $k_\text{F} \lesssim q^*$. In this situation, the electron-phonon interaction for flexural modes diverges and is usually cut off. Several attempts have been made to systematically account for the anharmonic effects in the transport properties of graphene \cite{Mariani2010,Castro2010b,Ochoa2011,Gorniy}. However, only little attention to this problem has been given in the context of conventional semiconductors \cite{Fischetti}. Despite the availability of advanced \emph{ab initio} computational techniques developed to describe electron-phonon scattering \cite{Marzari,Giustino,Ponce,Sohier}, their applicability is limited with respect to the above-mentioned effects.

Group V elemental semiconductors is an emerging class of 2D materials with attractive electronic properties \cite{Sofer,Ares2018}. Hexagonal (A7) 2D phases of phosphorus (P), arsenic (As), and antimony (Sb) are among the most recent materials fabricated experimentally \cite{blueP-syn,Sb-syn,As-syn}.
Apart from being mechanically stable \cite{blueP,AsSb_stability}, they are proposed to have excellent electron transport characteristics \cite{Wang,AsSb_mobility} and a high degree of gate tunability \cite{Ghosh,Shu,Prishchenko} due to the buckled crystal structure. Mechanically, these materials are highly flexible with the flexural rigidities in the range 0.3--0.8 eV \cite{blueP,Lugovskoi}, which is significantly smaller compared to graphene \cite{kappaC}. This property allows us to expect a considerable anharmonic coupling in these materials, and its possible influence on the electron transport properties.

In this paper, we focus on a theory for phonon-mediated charge carrier scattering in 2D semiconductors taking the coupling between the acoustic phonons into account. We use the diagrammatic approach to calculate the anharmonic contribution to the electron-phonon scattering rate, ending up at two plausible approximations. We apply the developed theory to calculate electron and hole mobilities in three representative group V isotropic semiconductors: hexagonal monolayers P (also known as blue phosphorene), As (arsenene), and Sb (antimonene). To this end, we estimate relevant model parameters from first-principles density functional theory (DFT) calculations. We find that although flexural phonons tend to reduce the mobility at concentrations below 10$^{14}$ cm$^{-2}$ by 15--30\%, their effect reaches its maximum at around 10$^{12}$ cm$^{-2}$. At lower dopings, the mobility becomes essentially independent of the concentration, being the manifestation of the anharmonic coupling between the flexural and in-plane phonon modes.

The paper is organized as follows. In Sec.~\ref{sec2}, we present a theory of charge carrier scattering in 2D semiconductors, explicitly considering in-plane and out-of-plane phonons, as well as their anharmonic coupling. In Sec.~\ref{sec3}, we use first-principles calculations to estimate model parameters for hexagonal P, As, and Sb monolayers. The results of numerical calculations are presented in Sec.~\ref{sec4}, where we investigate the interplay between in-plane and flexural phonons in the context of charge carrier scattering, as well as estimate upper limits for the carrier mobility in the materials under consideration. In Sec.~\ref{sec5}, we summarize our results, and conclude the paper.

\begin{section}{Theory}{\label{sec2}}

\subsection{Carrier mobility and relaxation time}
Carrier mobility of isotropic electron gas can be expressed as $\mu_c=\sigma/ne$,
where $n$ is the carrier concentration and $\sigma$ is the dc conductivity, which has the form \cite{Ziman}
\begin{equation}
    \sigma = \frac{e^2}{2S}\sum_{\bf k} \tau_{\bf k} v_{\bf k}^2 \left( -\frac{\partial f(\varepsilon_{\bf k})}{\partial \varepsilon_{\bf k}}  \right).
    \label{cond1}
\end{equation}
Here, $S$ is the sample area, $\varepsilon_{\bf k}$ is the band energy, $v_{\bf k}$ is the group velocity, $\tau_{\bf k}$ is the momentum-dependent scattering relaxation time, and $f(\varepsilon_{\bf k})=\{1+\mathrm{exp}[(\varepsilon_{\bf k}-\mu)/T])\}^{-1}$ is the Fermi-Dirac distribution function.
The energy dispersion of charge carriers is assumed to have the following form near the band edge:
\begin{equation}
\varepsilon_{\bf k}=\varepsilon_0 + \frac{\hbar^2({\bf k}-{\bf k}_0)^2}{2m},
\label{efmassfit}
\end{equation}
where $m$ is the electron ($m^\mathrm{e}$) or hole ($m^\mathrm{h}$) effective mass and $\varepsilon_0$ and ${\bf k}_0$ are the energy and wave vector determining the position of band edges, respectively.
The chemical potential $\mu$ can be determined from the carrier concentration $n$ as
\begin{equation}
\mu = T\,\mathrm{ln}\left[ \mathrm{exp}\left({\frac{2\pi\hbar^2n}{g_\text{v}g_\text{s}mT}}\right) - 1 \right],
\end{equation}
where $g_\text{v}g_\text{s}$ is the valley-times-spin-degeneracy factor.
In turn, one can recast Eq.~(\ref{cond1}) in the form,
\begin{equation}
\sigma = \frac{g_\text{v}g_\text{s}e^2\tau}{4\pi\hbar^2}\left\{\mu+2T\,\mathrm{ln}\left[2\mathrm{cosh}\left(\frac{\mu}{2T}\right)\right] \right\},
\label{sigma2}
\end{equation}
which in the limit $T\rightarrow 0$ transforms into the standard conductivity formula $\sigma=ne^2\tau/m$.
{
Here we assume the relaxation time to be averaged over thermally broadened Fermi surface states $\tau \equiv \langle \langle \tau_{\bf k} \rangle \rangle_{\mathrm{FS}}$, which is dependent on the chemical potential $\mu$ and temperature $T$.
}
The corresponding expression for $\tau$ can be obtained from the variational solution of the isotropic Boltzmann transport equation \cite{Ziman}, yielding
\begin{equation}
    \frac{1}{\tau} = \frac{2\pi}{\hbar S} \frac{\sum_{{\bf k}{\bf k}'} \left(-\frac{\partial f(\varepsilon_{\bf k})}{\partial \varepsilon_{\bf k}}\right) \delta(\varepsilon_{\bf k}-\varepsilon_{{\bf k}'}) (1-\mathrm{cos}\theta_{{\bf k}{\bf k}'}) \langle|V_{{\bf k}{\bf k}'}|^2\rangle  }   {\sum_{\bf k} \left(-\frac{\partial f(\varepsilon_{\bf k})}{\partial \varepsilon_{\bf k}}\right)},
    \label{tau1}
\end{equation}
where ${\bf k}={\bf k}'+{\bf q}$ with ${\bf q}$ being the scattering wave vector, 
{$V_{{\bf k}{\bf k}'}$ is the effective electron-phonon scattering potential defined in Sec.~\ref{2b}, and $\langle ...\rangle=\mathrm{Tr}(e^{-\beta H}...)/\mathrm{Tr}(e^{-\beta H})$ with $\beta=1/T$ denotes thermal averaging over the phonon states with the Hamiltonian $H$ defined below.}
In Eq.~(\ref{tau1}), emission and absorption of phonons are not explicitly involved in the energy conservation law, which assumes elastic scattering, i.e., $\hbar \omega_{k_\text{F}} \ll \varepsilon_F$ with $\omega_{k_\text{F}}$ being the phonon frequency at the Fermi wave vector. This condition is satisfied for acoustic phonons considered in our study.
After some algebra, Eq.~(\ref{tau1}) can be rewritten in a more convenient form as follows:
\begin{equation}
\label{tau2}
\tau^{-1}=\frac{m}{2\pi \hbar^3f(0)} \int_{0}^{\infty}d\varepsilon \left( -\frac{\partial f(\varepsilon)}{\partial \varepsilon}  \right) \int_0^{2k}\frac{dq~q^2 \langle|V_{q}|^2\rangle}{k^2\sqrt{k^2-q^2/4}},
\end{equation}
where $k=\sqrt{ 2m\varepsilon}/ \hbar$ plays the role of the Fermi wave vector at $\mu \gg T$.
{
We note that Eq.~(\ref{tau2}) is only valid for conventional 2D semiconductors with quadratic dispersion of charge carriers. Deviations from the quadratic dispersion, as well as the presence of a dielectric environment (e.g., substrates), could result in a considerable modification of the energy dependence of the scattering rate (see, e.g., Refs.~\onlinecite{Hwang,Kaasbjerg12,Min}).
}

\subsection{Electron-phonon scattering}
\label{2b}
Here we first define the scattering potential.
We restrict ourselves to the long-wavelength limit and take both in-plane and out-of-plane (flexural) phonon modes into account. We assume that electrons interact with phonons through the deformation potential of the form
\begin{equation}
V({\bf r})=g\,u_{\alpha\alpha}({\bf r}),    
\label{Vr}
\end{equation}
where 
\begin{equation}
u_{\alpha\beta}({\bf r})=\frac{1}{2}\left[\partial_{\alpha}u_{\beta}({\bf r})+\partial_{\beta}u_{\alpha}({\bf r})+\partial_{\alpha}h({\bf r})\partial_{\beta}h({\bf r})\right]
\label{uaa}
\end{equation}
is the strain tensor and $g$ is the coupling constant, whereas $u_{\alpha}({\bf r})$ and $h({\bf r})$ are in-plane and out-of-plane displacement fields, respectively.

{
The form of the deformation potential in Eq.~(\ref{Vr}) is justified by the following considerations. In the absence of substrates or external tension, the electronic structure is invariant under rotations of the crystal as a whole in three-dimensional space~\cite{guinea_collective_2014}. Therefore, $V\left({\bf r}\right)$ must be invariant under the following transformations:
\begin{equation}
\begin{split}
& u_{\alpha}\left({\bf r}\right) \rightarrow u_{\alpha}({\bf r}) - \delta \varphi_{\alpha} h({\bf r})\\
& h({\bf r}) \rightarrow h({\bf r}) + \delta \varphi_{\alpha}\left(r_{\alpha} + u_{\alpha}({\bf r})\right)~.\\
\end{split}
\label{rotation}
\end{equation}
To lowest order in $\partial_{\alpha} u_{\beta}\left({\bf r}\right)$ and $\partial_{\alpha} h\left({\bf r}\right)$, this implies that $V({\bf r})$ can only depend on the components of the strain tensor, which are scalar quantities under the transformation in Eq.~(\ref{rotation}) \cite{footnote}. 
In principle, $V({\bf r})$ can depend on all components of $u_{\alpha \beta}\left({\bf r}\right)$. Here we assume the isotropic form in Eq.~(\ref{Vr}) for simplicity. Notice that, as a consequence of rotational invariance, in-plane and out-of-plane coupling constants in Eq.~(\ref{Vr}) are not independent~\cite{SanJose, guinea_collective_2014}.
}

{
In view of the absence of horizontal mirror ($\sigma_\mathrm{h}$) symmetry in buckled hexagonal semiconductors ($D_{3\mathrm{d}}$ point group) being studied in this work, Eqs.~(\ref{Vr}) and (\ref{uaa}) require additional justification. Particularly, as has been shown in Ref.~\onlinecite{Fischetti}, single-phonon processes associated with flexural modes are highly relevant for non-mirror-symmetric 2D Dirac materials like silicene and germanene. Such processes correspond to additional terms in Eq.~(\ref{Vr}) that are linear in out-of-plane displacements, i.e., $\nabla h$. However, in our case those terms are irrelevant for the following reasons. Here we are focused on lightly doped semiconductors with quadratic energy dispersion represented by Eq.~(\ref{efmassfit}). Moreover, the intervalley scattering can be neglected because $k_F\ll k_0$. Under these assumptions, it is obvious from in-plane symmetry that $\nabla h$ terms can only enter the interacting potential via the transformation ${\bf k}_0 \rightarrow {\bf k}_0+\alpha \nabla h$.  
However, such terms are forbidden by rotational symmetry in three-dimensional space. 
This consideration does not forbid the appearance of higher-gradient terms in the deformation potential. As can be shown, $\nabla^2 h$ terms are forbidden, too, which follows from the combination of inversion and time-reversal symmetry, particularly because ${\bf k}_0\rightarrow -{\bf k}_0$. On the other hand, $\nabla^3 h$ and higher terms acquire additional power of phonon wave vector ${\bf q}$ in comparison to $\nabla u$ terms, and since $|{\bf q}| \lesssim 2k_F$, such terms have smallness in electron concentration and, therefore, can be neglected.  
}

Let us now evaluate the scattering probability $\langle |V_{\bf q}|^2 \rangle$. 
The Fourier transform of the scattering potential $V({\bf r})$ is then given by
\begin{equation}
V_{\bf q}=igq_{\alpha}u_{\alpha}({\bf q})+\frac{g}{2}f_{\alpha \alpha}({\bf q}),
\label{Vq}
\end{equation}
where $u_{\alpha}({\bf q})$ are the Fourier components of the in-plane displacement field $u_{\alpha}({\bf r})$ and $f_{\alpha \beta}({\bf q})=- S^{-\frac{1}{2}}\sum_{\bf k}k_{\alpha}(q_{\beta}-k_{\beta})h_{\bf k}h_{{\bf q}-{\bf k}}$ is the Fourier transform of $f_{\alpha \beta}({\bf r})=\partial_{\alpha}h({\bf r})\partial_{\beta} h({\bf r})$. In the long-wavelength limit,
 fluctuations of the displacement fields are described by the Hamiltonian~\cite{Nelson, Nelson-orig, le_doussal_self-consistent_1992}
\begin{equation}
    H=\frac{1}{2}\int d{\bf r}\left\{ \kappa [\nabla^2h({\bf r})]^2 +\lambda u^2_{\alpha \alpha}({\bf r}) +  2 G u^2_{\alpha \beta}({\bf r}) \right\},
    \label{hphon}
\end{equation}
where $\lambda$ and $G$ are the Lam\'{e} parameters. 
We assume that fluctuations of the displacement fields can be described classically. Neglecting quantum effects is fully justified at not-too-small temperatures, when $ \hbar \omega (k_F) \ll T$. In the calculation of the thermal average $ \langle \left|V_{{\bf q}} \right|^{2}\rangle$, we can take advantage of the fact that the Hamiltonian in Eq.~(\ref{hphon}) is quadratic with respect to the in-plane displacement fields $u_{\alpha}\left({\bf r}\right)$, which allows one to integrate them out exactly~\cite{Nelson, Nelson-orig, le_doussal_self-consistent_1992}. After some algebraic manipulations~\cite{Nelson, Nelson-orig}, we obtain the following expression for the scattering matrix {(see the Appendix for details)}:
\begin{equation}
   \langle |V_{\bf q}|^2 \rangle = g^2\frac{T}{Y}(1-\nu^2)+g^2\frac{(1-\nu)^2}{4}\langle f^T({\bf q})f^T(-{\bf q})\rangle,
   \label{Vq2}
\end{equation}
where $\nu=\lambda/(\lambda+2G)$ is the 2D Poisson ratio and  $Y= \lambda (1-\nu^2)/\nu$.
{
Being focused on the processes with small momentum transfer ${\bf q}={\bf k}-{\bf k}'$, we ignored the effect of the wave-function overlap between the initial ${\bf k}$ and final ${\bf k}'$ states, which might lead to some overestimation of the scattering probability. 
}
In Eq.~(\ref{Vq2}), the first term coincides with the usual contribution of in-plane phonons in the harmonic approximation, whereas the second term describes both out-of-plane and cross interactions generated by anharmonic coupling; 
$f^{T}\left({\bf q}\right) \equiv  \left(\delta_{\alpha \beta} - q_{\alpha} q_{\beta}/q^{2}\right) f_{\alpha \beta}\left({\bf q}\right)$ denotes the transverse part of $f_{\alpha \beta}\left({\bf q}\right)$. In what follows, we use two different approximations to calculate the correlation function $\langle f^{T}\left({\bf q}\right) f^{T}\left(-{\bf q}\right)\rangle$. 

\emph{Wick decoupling approximation}. In the first approximation, we assume the validity of the Wick theorem, which yields $\langle h_{\bf k} h_{{\bf q}-{\bf k}}h_{{\bf k}'}h_{-{\bf q}-{\bf k}'} \rangle \simeq \left( \delta_{{\bf k},{{\bf q}+{\bf k}'}} + \delta_{{\bf k}, -{\bf k}'}\right)\langle h_{\bf k} h_{-\bf k}\rangle \langle h_{{\bf k}-{\bf q}}h_{{\bf q}-{\bf k}} \rangle$~\cite{Katsnelson-Book}. In terms of diagrams, this assumption corresponds to the neglect of vertex contributions to the two-particle Green's function. In this approximation, the desired correlation function reads:
\begin{equation}
\begin{split}
 \langle f^{T}({\bf q}) & f^{T}(-{\bf q})\rangle  \simeq  \Pi\left({\bf q}\right) \\  & \equiv 2\int \frac{d^2{\bf k}}{(2\pi)^2} \left[\frac{({\bf q}\times{\bf k})^2}{q^2}\right]^{2} G({\bf k})G({\bf q}-{\bf k})~,
\label{pi}
\end{split}
\end{equation}
where $G({\bf q})\equiv \langle h_{\bf q}h_{-\bf q}\rangle$ is the single-particle correlation function of out-of-plane fields. The function $\Pi\left({\bf q}\right)$ exhibits a crossover from a harmonic behavior for $q \gg q^{*}$ to an anomalous scaling behavior for $q \ll q^{*}$, which is controlled by anharmonic interactions. In this work, we addressed the calculation of $\Pi\left({\bf q}\right)$ within the self-consistent screening approximation (SCSA)~\cite{le_doussal_self-consistent_1992, le_doussal_anomalous_2018}. In the limit of large and small wave vectors, $\Pi \left({\bf q}\right)$ behaves asymptotically as:
\begin{equation}
      \Pi(\q) = 
          \begin{cases}
               \frac{3}{8\pi}\frac{T^2}{\kappa^{2} q^2}    &\text{, } q\gg q^* \\
               C \left( \frac{T}{\kappa} \right)^{2-\eta}\left(\frac{\kappa}{Y} \right)^{\eta} q^{-2+2\eta}     &\text{, } q\ll q^* 
          \end{cases},
          \label{pi_asy}
\end{equation}
where $\eta$ is an universal exponent and $C$ is a dimensionless constant. The value of $\eta$ is close to $0.821$ within the SCSA, in the physical case of two-dimensional membranes embedded in three-dimensional space. Through a numerical solution of the SCSA equations~\cite{le_doussal_self-consistent_1992, le_doussal_anomalous_2018}, we determined $\Pi \left({\bf q}\right)$ for arbitrary values of the wave vector, including the intermediate scale $q \simeq q^{*}$ and we obtained the value $C \simeq 2.60$ for the dimensionless amplitude in Eq.~(\ref{pi_asy}).

\emph{Screening approximation}. In the second approximation, we consider the following expression for the correlation function:
\begin{equation}
 \langle f^T({\bf q})f^T(-{\bf q})\rangle \simeq \frac{\Pi(\bf q)}{1+\frac{Y}{4T}\Pi({\bf q})}.
 \label{screening_expression}
\end{equation}
Equation~(\ref{screening_expression}) can be derived diagrammatically by summing a geometric series of polarization bubble diagrams connected by bare interaction lines, the same diagrams which determine, within the SCSA, the screened interaction between flexural phonons~\cite{le_doussal_self-consistent_1992, le_doussal_anomalous_2018}. An expression equivalent to  Eq.~(\ref{screening_expression}) is reported in Ref.~\cite{guinea_collective_2014}.

Adopting Eq.~(\ref{screening_expression}), one arrives at the following asymptotic limits for the correlation function:
\begin{equation}
      \langle f^T({\bf q})f^T(-{\bf q})\rangle = 
          \begin{cases}
               \frac{3}{8\pi}\frac{T^2}{\kappa^2 q^2}    & \text{, }q\gg q^* \\
               \frac{4T}{Y} - \frac{16}{C}\frac{T^2}{\kappa^2} \left( \frac{TY}{\kappa^2} \right)^{-2+\eta} q^{2-2\eta}     &\text{, } q\ll q^* 
          \end{cases}
          \notag
\end{equation}
As expected, the two approximations result in the same behavior at $q\gg q^*$, while the result at $q\ll q^*$ is different. In the Wick decoupling approximation, the electron-phonon coupling $\langle|V_{\bf q}|^2\rangle$ scales as $1/q^{2-2\eta}$ for small $q$ and it diverges in the limit $q \rightarrow 0$, although the divergence is slow, because the exponent $\eta$ is not very different from $1$. By contrast, in the screening approximation $\langle|V_{\bf q}|^2\rangle$ approaches the constant value $ 2g^2\frac{T}{Y}(1-\nu)$ for $q \rightarrow 0$. 
This expression ensures that, in the screening approximation, scattering on in-plane phonons always gives a larger contribution than scattering on out-of-plane phonons. In the discussion below, the two approximations will be considered as the upper and lower limits for the relaxation time.

The expressions presented above allow one to estimate the dc conductivity (or mobility) of isotropic 2D electron gas in the presence of elastic scattering on acoustic phonons. Importantly, this approach is equally valid for small carrier concentrations, where scattering involves the coupling between phonons modes.
\end{section}

\section{Estimation of material constants}{\label{sec3}}

\subsection{Calculation details}

To estimate material constants, we use first-principles DFT calculations. All calculations have been carried out using the projected augmented-wave \cite{Blochl} 
formalism as implemented in the Vienna \emph{ab initio} simulation package 
({\sc vasp}) \cite{Kresse1996a,Kresse1996b,Kresse1999}. 
To describe exchange-correlation
effects, we employed the gradient-corrected approximation in the parametrization of
Perdew-Burke-Ernzerhof \cite{pbe}.
An energy cutoff of 500 eV for the plane waves and the
convergence threshold of 10$^{-9}$ eV were used in all cases. To avoid spurious interactions 
between the cells, a vacuum slab of 50~\AA~ was added in the direction perpendicular to
the 2D sheet. 
Structural relaxation including the optimization of in-plane lattice constants 
was performed until the forces acting on atoms were less than 10$^{-3}$ eV/\AA. Unlike the cases of P and As monolayers, spin-orbit interaction in monolayer Sb is important \cite{Rudenko2017} and, therefore, has been taken into account.

Hexagonal monolayers studied in this work adopt honeycomb structure with vertically displaced (buckled) sublattices. We obtained the following optimized values of the lattice constant ($a$) and buckling parameter: 3.28~\AA~and 1.24~\AA~for P, 3.61~\AA~and 1.40~\AA~for As,  4.12~\AA~and 1.64~\AA~for Sb, respectively. Primitive hexagonal cells were used in most of the calculations, for which a Monkhorst-Pack~\cite{kmesh} $\Gamma$-centered (32$\times$32) mesh was adopted to sample the Brillouin zone. To calculate the Poisson ratio and Young modulus, a rectangular $(\sqrt{3}/2 \times 1)a$ unit cell was used, together with a $(32\times24)$ {\bf k}-point mesh.  
To induce flexural deformations needed for the calculation of flexural rigidities, we considered rectangular (6$\sqrt{3}\times$1)$a$ supercells, such that the deformation period amounts to $l=6\sqrt{3}a\sim40$~\AA~ for each system considered.
We first apply a field of 1D sinusoidal out-of-plane deformations in the form $h(x,y)=h \, \mathrm{sin}(qx)$ to each system, where $h$ the deformation amplitude and $q=2\pi/l$. We then perform full structural relaxation for a series of fixed amplitudes $h$, which allows us to obtain the elastic energies. In the calculations with rectangular supercells, a ($4\times24$) {\bf k}-point mesh was used.

\subsection{Effective masses}

\begin{figure}[tp]
\includegraphics[width=0.99\columnwidth]{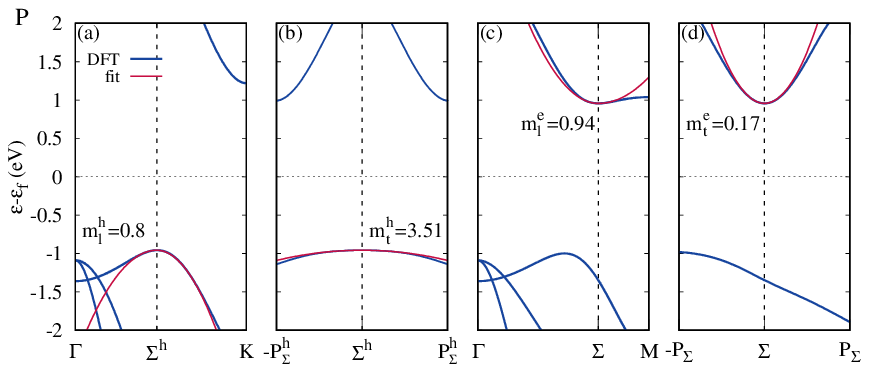} 
\vspace*{0.1cm}
\includegraphics[width=0.49\columnwidth]{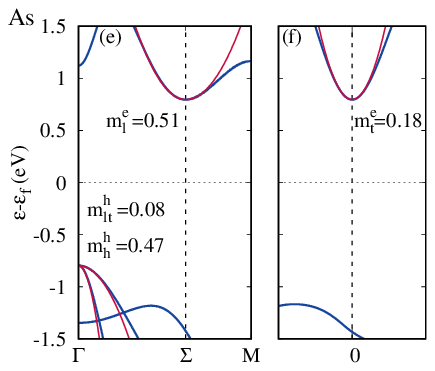}
\includegraphics[width=0.49\columnwidth]{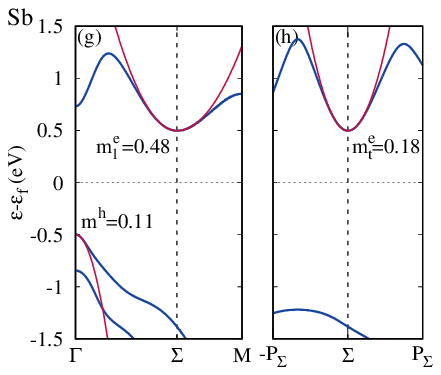}
	\caption{(Color online) Blue: DFT band structure calculated along high-symmetry directions of the Brillouin zone for P [(a)--(d)], As [(e) and (f)], and Sb [(g) and (h)] monolayers. Red:
	Band edges fitted within the effective mass approximation [Eq.~(\ref{efmassfit})].
	Due to the degeneracy of VBM in case of As (g), two values are given for $m^{\text{h}}$, corresponding to light ($m^\text{h}_\text{lt}$) and heavy ($m^\text{h}_\text{h}$)  holes. For all three systems considered, two types of electron effective masses are provided:
	longitudinal ($m_{\text{l}}$) and transverse ($m_{\text{t}}$). The same distinction applies to the hole effective mass of P monolayer [(a) and (b)]. 
	}
    \label{fig:masses}
\end{figure}

Electron and hole effective masses are estimated from the band structures by fitting the conduction band minimum (CBM) and valence band maximum (VBM), respectively, using the expression given by Eq.~(\ref{efmassfit}).
The calculated band structure used for the fitting, and the corresponding effective masses are given in Fig.~\ref{fig:masses}. 
In all three systems considered, CBM resides at the $\Sigma$ point along the $\Gamma$--M high symmetry line. The corresponding bands are spin ($g_\text{s}=2$) and valley degenerate ($g_\text{v}=2$). The constant-energy (Fermi) contours at small energies form elliptical pockets \cite{Lugovskoi}, meaning that the electron effective mass is different in the direction perpendicular to $\Gamma$--M, denoted in Fig.~\ref{fig:masses} as $-$P$_{\Sigma}$--P$_{\Sigma}$. We denote the corresponding masses as longitudinal ($m_\text{l}^\text{e}$) and transverse ($m^\text{e}_\text{t}$), as shown in Fig.~\ref{fig:masses}. In practical calculations, we use geometrically averaged effective masses, $m^\text{e}=\sqrt{m^\text{e}_\text{l}m^\text{e}_\text{t}}$, with the resulting values listed in Table~\ref{constants}. 
The obtained values overall agree with the literature data~\cite{AsSb_mobility2015,AsSb_mobility,AsSb_mobility2017}.

The case of holes in monolayer P is similar to the case of electrons. Since VBM is located at the $\Sigma^\text{h}$ point along $\Gamma$--K symmetry line, the effective mass tensor is anisotropic, and $m^{\text{h}}$ can be calculated as the average of its longitudinal and transverse components.
The situation with the hole effective masses in As and Sb is different.
In these systems, VBM is located at the $\Gamma$ point, making the Fermi contour isotropic with the direction-independent hole effective masses. Unlike monolayer Sb, where VBM at the $\Gamma$ point is doubly degenerate ($g_\text{s}=2$), VBM of monolayer As is four times degenerate due to the absence of strong spin-orbit coupling. As one can see, there are two different bands forming the valence band in monolayer As, meaning that there are two transport channels for holes in As. To take them into account, specifically for As we define the transport effective mass, which is the sum of light and heavy hole masses, $m^\text{h}_\text{tr}=m^\text{h}_\text{lt} + m^\text{h}_\text{h}$. On the other hand, the density of states effective mass is different, $m^\text{h}_{\text{d}}=\sqrt{m^\text{h}_\text{h}m^\text{h}_\text{lt}}$. Given that the conventional electron-phonon contribution to the mobility can be expressed through the product $m_\text{d}m_\text{tr}$ \cite{Takagi}, we employ $m^\text{h}=\sqrt{m^\text{h}_{\text{d}}m^\text{h}_{\text{tr}}}$ in the calculations of As. In all other systems, $m_\text{d}$ and $m_\text{tr}$ are equivalent.

\begin{table}[t]
	\caption{Relevant material constants estimated from first principles for group V hexagonal (A7 phase) 2D semiconductors: Electron (hole) effective mass $m^{\text{e}(\text{h})}$ (in units of free electron mass), for the case of hole doped arsenene averaged effective mass is given, electron (hole) deformation potential $g^{\text{e}(\text{h})}$ (in eV), Young modulus $Y$ (in eV$\cdot$\AA$^{-2}$), Poisson ratio $\nu$, flexural rigidity $\kappa$ (in eV), and a characteristic wave vector $q^*$ at $T=$~300~K (in \AA$^{-1}$).}
    \label{constants}
	\begin{ruledtabular}
		\renewcommand\arraystretch{1.3}
		\renewcommand\tabcolsep{1.1pt}
		\begin{tabular}{lcccccccc}
System &  $m^\text{e}$ & $m^\text{h}$ &  $g^\text{e}$ &  $g^\text{h}$ &  $Y$  & $\nu$ & $\kappa$ & $q^*$ \\ \colrule
P      &  0.40  & 1.72  &  2.26  &   1.34   &  4.53 &  0.11 &  0.70    &  0.12 \\
As     &  0.30  &  \phantom{$^*$}0.33$^{*}$  &  1.34  &   6.13   &  3.07 &  0.16 &  0.51    &  0.14 \\
Sb     &  0.29  & 0.11  &  0.86  &   5.99   &  1.86 &  0.19 &  0.33    &  0.17 \\
	\end{tabular}
	\end{ruledtabular}
\raggedright{$^{*}m=\sqrt{m_\text{d}m_\text{tr}}$}, see Sec.~\ref{sec3}~B for details.
\end{table}
\vspace*{-0.6cm}

\subsection{Elastic properties}

\subsubsection{Flexural rigidities}

Flexural rigidities $\kappa$ can be estimated from first principles starting from a macroscopic expression for the elastic energy density of a 2D membrane, Eq.~(\ref{hphon}). In the absence of strain, it reads
\begin{equation}
E_{\mathrm{elas}}=\frac{\kappa}{2} \! \int \! d{\bf r} \, (\nabla ^2 h)^2.
\label{elastic}
\end{equation}
The elastic energy $E_{\mathrm{elas}}$ can be directly obtained from DFT calculations as the total energy difference $\Delta E_{\mathrm{tot}}$
between the corrugated and flat states of a membrane, $E_{\mathrm{elas}}=\Delta E_{\mathrm{tot}}$. 
By employing the sinusoidal height field in the form $h(x,y)=h \, \mathrm{sin}(qx)$, the constant $\kappa$ can be estimated by a regression fit from
\begin{equation}
\Delta E_{\mathrm{tot}}=\frac{\kappa}{4}\frac{1}{R^2},
\label{kapfit}
\end{equation}
where $R$ is the curvature radius of a sine wave at its extrema, 
given by $1/R=hq^2$.

In Fig.~\ref{DFT}(b), we show the calculated dependencies $\Delta E_{\mathrm{tot}}$ as a function of $hq^2$, which allow us to
estimate $\kappa$ for the systems under consideration (see Table \ref{constants} for summary). In all cases the expression given by
Eq.~(\ref{kapfit}) perfectly fits DFT data.

\subsubsection{In-plane elastic constants}
In the absence of flexural deformations, the elastic energy density $E_{\mathrm{elas}}$ 
of a hexagonal strained 2D material can be expressed from Eq.~(\ref{hphon}) as
\begin{equation}
E_{\mathrm{elas}} = \frac{Y\nu}{2(1-\nu^{2})} u_{\alpha\alpha}^2 + \frac{Y}{2(1+\nu)} u^2_{\alpha \beta}.
\label{eq:eelast}
\end{equation}
The Young modulus $Y$ can be found straightforwardly from DFT calculations by applying a uniaxial deformation in the desired direction and allowing full relaxation in other directions:
\begin{equation}
    E_{\mathrm{elas}}=\frac{1}{2}Y u_{\mathrm{\alpha\alpha}}^2\bigg\vert_{\sigma_{\beta\beta}=0},
\end{equation}
where $\sigma_{\beta\beta}$ is the component of the stress tensor in the relaxed direction.
Similarly, the Poisson ratio can be obtained from the same set of calculations:
\begin{equation}
    \nu=-\frac{\Delta l_{\beta\beta}}{\Delta l_{\alpha\alpha}}\bigg\vert_{\sigma_{\beta\beta}=0},
\end{equation}
where $\Delta l_{\alpha\alpha}$ and $\Delta l_{\beta\beta}$ are the change of lattice constants in the strained and relaxed directions, respectively.

\begin{figure}[btp]
	\includegraphics[width=\columnwidth]{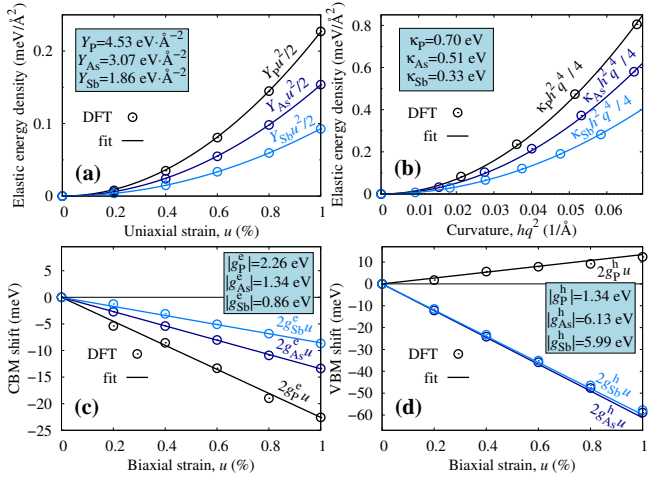}
	\caption{Elastic energies and band-edge shifts calculated as functions of in-plane and out-of-plane deformations for hexagonal monolayers P, As, and Sb.
Panels (a) and (b) are used to determine 2D Young modulus ($Y$) and flexural rigidity ($\kappa$), whereas panels (c) and (d) are used to determine electron and hole deformation potentials ($g^e$ and $g^h$). CBM and VBM stand for the conduction band minimum and valence band maximum, respectively. 
}
    \label{DFT}
\end{figure}

In Fig.~\ref{DFT}(a), we show the calculated $\Delta E_{\mathrm{tot}}$ as a function of uniaxial strain, which
perfectly fits the macroscopic energy expression given above. This allows us to readily estimate the Young modulus $Y$ for the materials under consideration. The calculated Poisson ratios are listed in Table \ref{constants}. 

\subsection{Deformation potentials}

\subsubsection{Interaction with in plane phonons}

To estimate the coupling of charge carriers with phonons from
DFT calculations, we adopt the concept of the deformation potentials \cite{BirPikus}. 
Assuming the long-wavelength limit, we quantify linear response of the band edges to external deformation.
To this end, we apply both tensile and compressive biaxial strain and 
obtain relative band shifts $\Delta E^{\mathrm{VBM}}$ and $\Delta E^{\mathrm{CBM}}$ as shown in Figs.~\ref{DFT}(b) and \ref{DFT}(c) as a function of strain $u=u_{\alpha\alpha}=u_{\beta\beta}$. The band shifts are calculated relative to the vacuum level $E_\text{vac}$, determined for every deformed state, that is $\Delta E=E-E_\text{vac}$, where $E$ is the energy of the corresponding band edge.

\begin{figure}[tp]
	\includegraphics[width=\columnwidth]{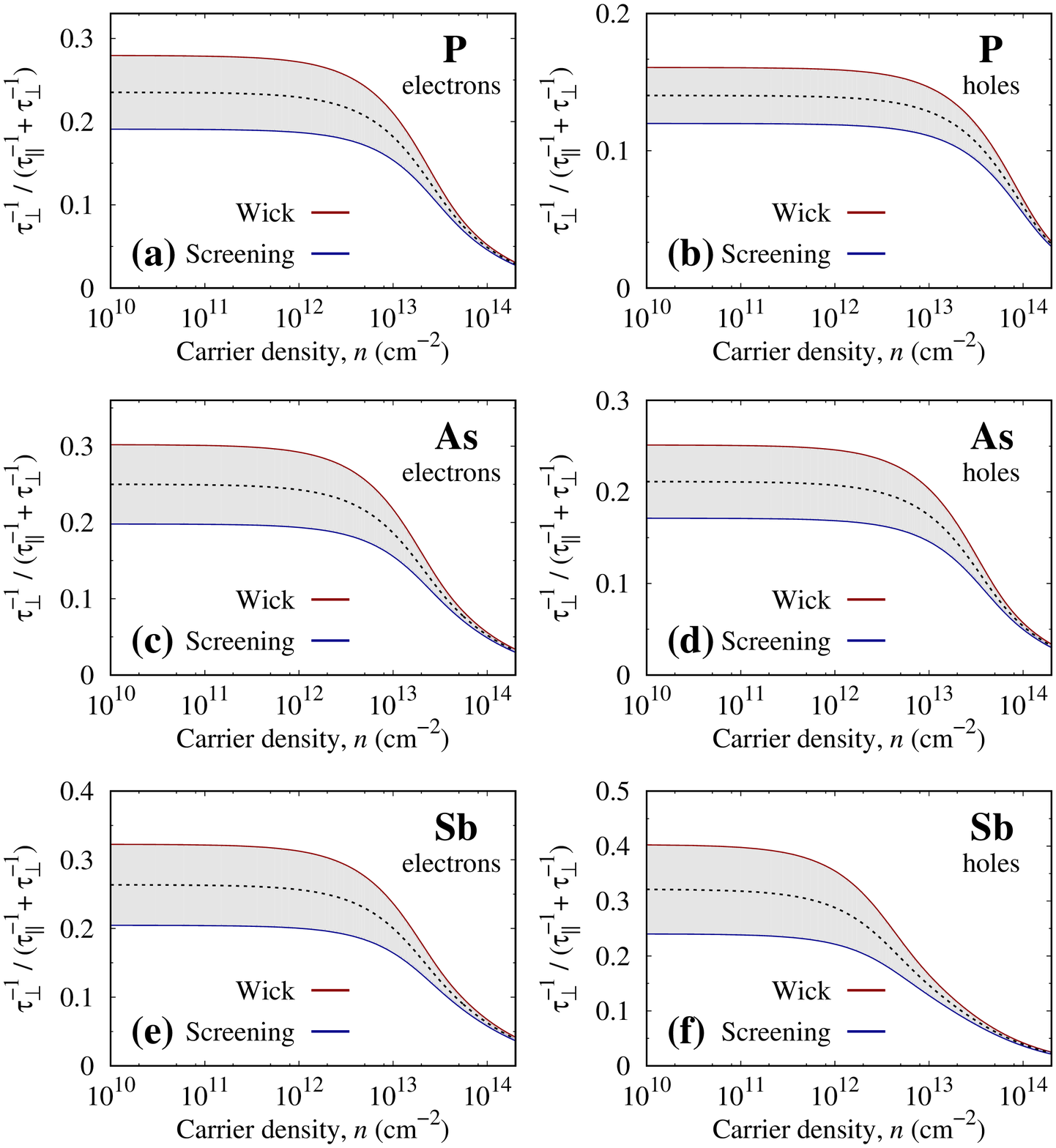}
	\caption{The ratio of carrier scattering rates induced by anharmonic flexural phonons ($\tau^{-1}_{\perp}$) to the total scattering rate ($\tau^{-1}_{\parallel}+\tau^{-1}_{\perp}$) calculated as a function of carrier concentration ($n$) for different types of dopings in monolayers  P, As, and Sb. Red and blue lines correspond to the Wick decoupling and screening approximations, respectively, as discussed in Sec.~\ref{sec2}. Dashed line represents the average of the two approximations with shaded area showing the uncertainty of the model. In all cases $T=300$ K.}
    \label{tautau}
\end{figure}

\begin{figure}[tp]
	\includegraphics[width=\columnwidth]{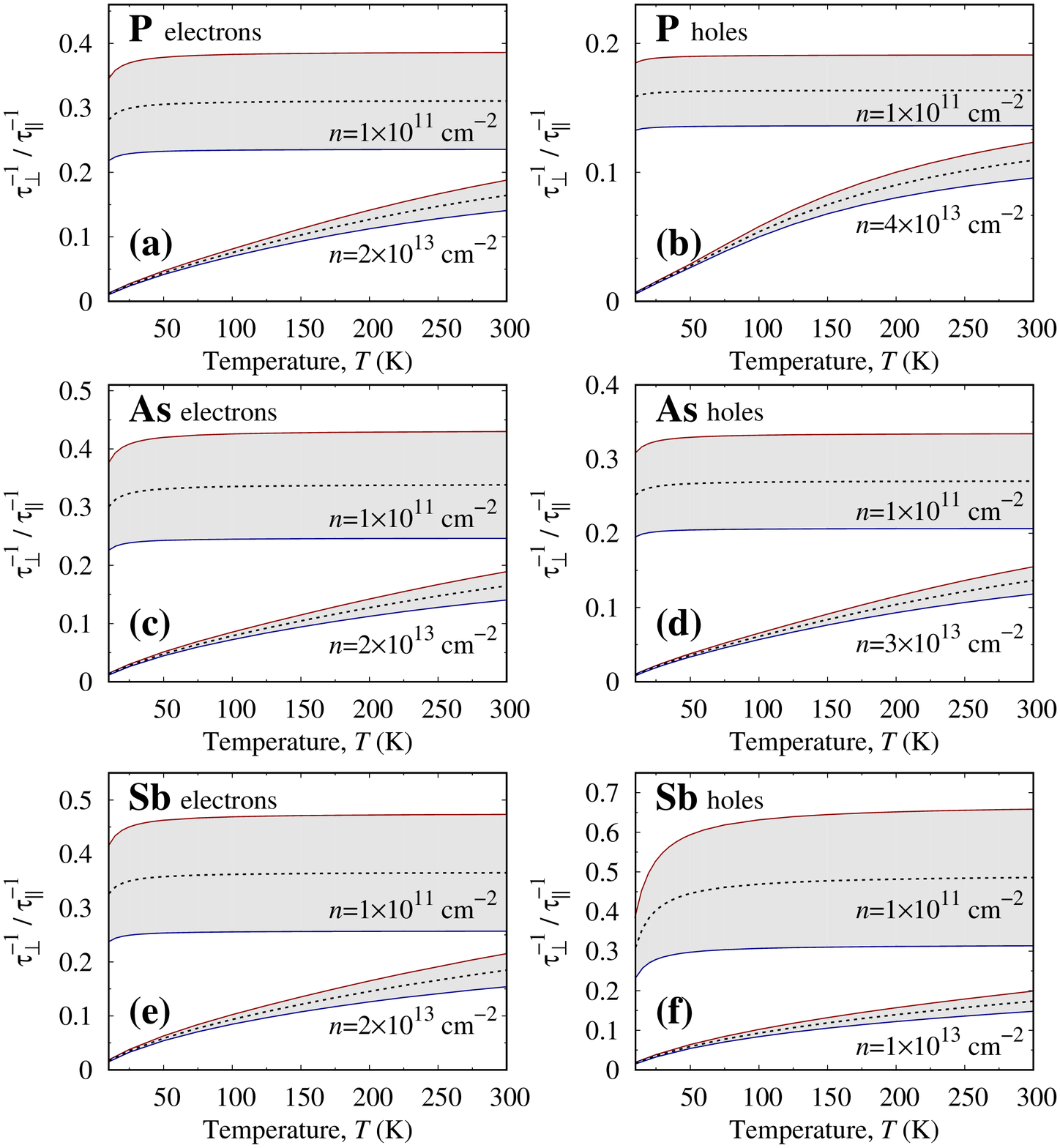}
	\caption{Temperature dependence of the ratio between flexural ($\tau^{-1}_{\perp}$) and in-plane ($\tau^{-1}_{\parallel}$) scattering rates calculated for different carrier concentrations ($n$) and doping types in P, As, and Sb monolayers. Red, blue, and dashed lines have the same meaning as in Fig.~\ref{tautau}.}
    \label{tautau_T}
\end{figure}

The interaction energy $E_{\mathrm{int}}$ is linear in strain and takes the form
\begin{equation}
E_{\mathrm{int}}=2g^{\mathrm{e(h)}}u
= \begin{cases}
\Delta E^{\mathrm{CBM}} \, \text{(electrons)} \\
\Delta E^{\mathrm{VBM}} \, \text{(holes).} \\
\end{cases}
\label{2cases_par}
\end{equation}
Since the sign of $E_{\mathrm{int}}$ depends on the strain type (tensile or compressive), 
the sign of $g^{\text{e}(\text{h})}$  is the matter of convention. 
The coupling constants are given in Fig.~\ref{DFT} and also summarized in Table \ref{constants}.
While the effective masses, as well as hole deformation potentials ($g^\text{h}$) for monolayers As and Sb agree well with the literature data, larger electron deformation potentials ($g^\text{e}$) have been reported for these materials  in~Refs.~\cite{AsSb_mobility2015,AsSb_mobility,AsSb_mobility2017}. The discrepancy can be attributed to a different approach utilized in those works. Particularly, uniaxial strain was used to quantify the band shifts, while we utilized biaxial strain in our work. 
{The latter approach implies that only diagonal elements of the deformation potential tensor are taken into account in our study [see Eq.~(\ref{Vr})]. This approximation is expected to quantitatively overestimate the mobilities calculated below.
}

\section{\label{sec4}Results and discussion}

We first analyze the role of flexural phonons and their anharmonic coupling in the scattering of charge carriers in 2D semiconductors P, As, and Sb. To this end, we decompose the total scattering rate into the two contributions $\tau^{-1}=\tau_{\parallel}^{-1}+\tau_{\perp}^{-1}$, each of which describes the scattering on pure in-plane modes [first term on the right-hand side of Eq.~(\ref{Vq2})] and on flexural modes subject to anharmonic coupling with in-plane modes [second term in the same equation]. In Fig.~\ref{tautau}, we show the dependence $\tau^{-1}_{\perp} / \tau^{-1}$ of the carrier concentration both for electron and hole dopings calculated at $T=300$ K. As long as the effective mass approximation holds, from Eq.~(\ref{tau2}) we have $\tau_{\parallel}^{-1}=mg^2T(1-\nu^2)/\hbar^3Y$, meaning that the doping dependence arises from the $\tau_{\perp}^{-1}$ term only. In the limit of large concentrations $\tau_{\perp}^{-1} / \tau^{-1} \rightarrow 0$ because $\tau_{\perp}^{-1} \sim 1/\varepsilon_F$, indicating that flexural phonons are negligible in this regime, which can be clearly seen from Fig.~\ref{tautau} for all the cases considered. At small concentration the behavior is quantitatively different for the two approximations describing the anharmonic coupling. Nevertheless, the trend is qualitatively similar: The contribution of flexural phonons becomes independent or weakly dependent on the doping. This behavior is consistent with ${\bf q}$-dependence of the scattering matrix $\langle|V_{\bf q}|^2\rangle$, which at $q \ll q^*$ behaves as $\sim$$q^{-0.36}$ within the Wick decoupling approximation and approaches a constant in the screening approximation. On average, the contribution of flexual phonons at room temperature does not exceed 30\% of the total scattering rate for all materials considered.

Let us now examine the temperature dependence of the scattering rate. As has been shown above $\tau_{\parallel}^{-1}\sim T$ independently of the carrier concentration. On the other hand, $\tau_{\perp}^{-1}\sim T^2$ in the harmonic regime ($q \gg q^*$), while it softens significantly in the strong anharmonic regime ($q \ll q^*$), yielding $\tau_{\perp}^{-1}\sim T^{1.18}$ or $\sim T$ depending on the approximation employed. The temperature dependence calculated for materials under consideration is shown in Fig.~\ref{tautau_T}. The trend is similar for all the cases considered. In particular, at low temperatures $\tau^{-1}_{\perp}/\tau^{-1}_{\parallel}$ is small as the electron gas is degenerate ($T \ll \varepsilon_F$), resulting in a negligible flexural phonon contribution. At higher temperatures, the behavior essentially depends on the carrier concentration. At small concentration around $10^{11}$ cm$^{-2}$, $\tau^{-1}_{\perp}/\tau^{-1}_{\parallel}$ is almost independent of the temperature, indicating the strongly anharmonic regime. At concentrations of the order of $10^{13}$ cm$^{-2}$, the corresponding ratio $\sim T$, as is expected from the limit $q \gg q^*$.

\begin{figure}[btp]
	\includegraphics[width=\columnwidth]{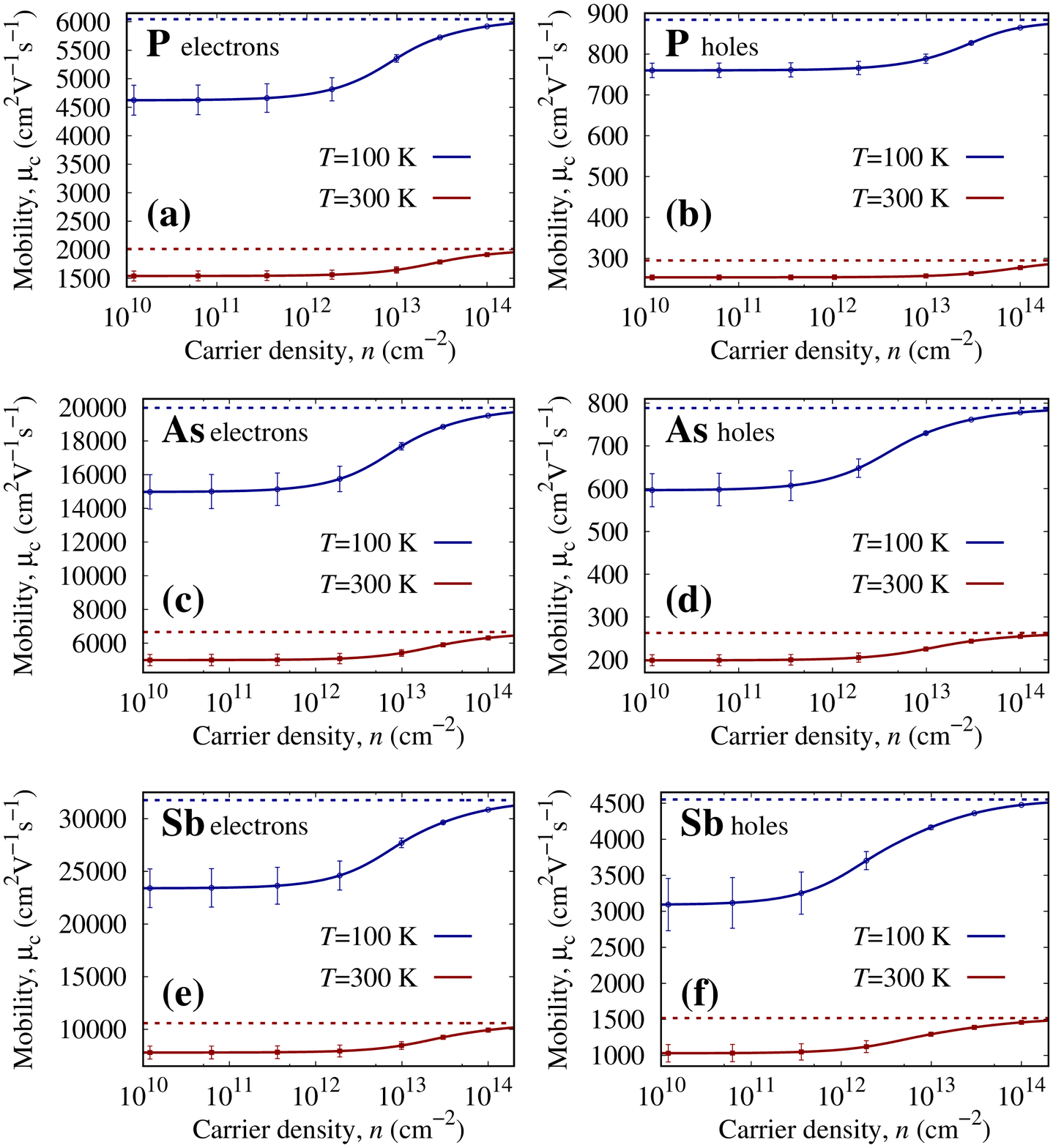}
	\caption{Carrier mobility calculated as a function of electron and hole doping in hexagonal monolayers P, As, and Sb for two different temperatures. Dashed lines correspond to the case of scattering by in-plane phonons only, while solid lines take both in-plane and flexural phonons into account. Error bars show uncertainty due to the approximations used to calculate the anharmonic coupling.}
    \label{mob}
\end{figure}

As a final step, we make a quantitative estimate of the carrier mobility, presented in Fig.~\ref{mob} for a wide range of electron and hole doping in P, As, and Sb. At large carrier concentrations, scattering by in-plane phonons dominates, and the mobility approaches its upper limit for a given temperature. As the concentration drops down, the mobility decreases by 15--30\% for different materials in the carrier density range from 10$^{12}$ to 10$^{14}$ cm$^{-2}$. This effect is attributed to the increased role of flexural phonons, whose scattering rate is inversely proportional to the carrier concentration. At lower dopings, the scattering behavior is modified by the strong anharmonic coupling between in-plane and flexural phonons. In particular, below 10$^{12}$ cm$^{-2}$ the mobility behavior is nearly independent of the material and doping type, resulting in a constant value.

Quantitatively, the mobilities are strongly dependent on the carrier type. The electron mobilities are significantly higher for all the systems considered, which can be attributed to relatively small effective masses and deformation potentials, listed in Table \ref{constants}. The lowest values not exceeding 1000 cm$^{2}$V$^{-1}$s$^{-1}$ are obtained for the hole doping of P and As, i.e., materials with relatively massive carriers. It is interesting to note that the electron-doped hexagonal (blue) phosphorus exhibits mobilities an order of magnitude larger than its orthorhombic (black) counterpart \cite{Rudenko2016}. 
It should be emphasized that the results presented in Fig.~\ref{mob} represent an upper limit for the intrinsic mobility. A number of important factors, such as, for instance, optical phonons, dielectric screening, many-body renormalization, and other effects, discussed, for example in Refs.~\onlinecite{Ponce} and \onlinecite{Fischetti2}, are not taken into account in the present study. This can explain some discrepancy with the results obtained from first principles \cite{AsSb_mobility,Wang,Sohier}. Moreover, under realistic conditions, charge carriers are subject to extrinsic scattering by impurities, defects, and substrates, which could further reduce the mobility by orders of magnitude.

\section{\label{sec5}Conclusion}
Flexural phonons are the inherent phenomena of 2D materials, which necessarily contributes to the electron-phonon scattering. Unlike conventional in-plane modes, whose contribution to electronic transport is weakly dependent on the carrier concentration, scattering on flexural phonons becomes more effective in the regime of low carrier densities. At the same time, this regime is characterized by the strong anharmonic coupling between in-plane and flexural modes, resulting in the modification of the scattering behavior. 

Here we developed a theory for the electron-phonon scattering in 2D semiconductors taking the anharmonic effects into account. We applied this theory to examine the scattering behavior in typical group V hexagonal monolayers. We showed that although the scattering rate is affected by the flexural phonons, the net effect does not exceed 30\%. Compared to the harmonic approximation, where the scattering by flexural phonons is divergent in the limit of low concentrations, the anharmonic coupling suppresses this effect significantly. For all materials considered, at carrier concentrations below 10$^{12}$ cm$^{-2}$ flexural phonon contribution turns out to be nearly independent of doping in a wide range of temperatures. Figure~\ref{mob} summarizes our findings.  

From the practical point of view, the suppression of flexural phonons (e.g., by a substrate or by strain) is considered to be one of the ways to enhance the mobility in 2D semiconductors. Our results show that this approach is highly limited and cannot lead to significant mobility gain. On the theory side, flexural phonons can be safely excluded from the consideration in the electronic transport calculations.

\begin{acknowledgments}
A.V.L. and M.I.K. acknowledge support from the research program ``Two-dimensional semiconductor crystals'' Project No. 14TWOD01, which is financed by the Netherlands Organisation for Scientific Research (NWO).
S.Y. acknowledges support from the National Key R\&D Program of China (Grant No. 2018FYA0305800).
Numerical calculations presented in this paper have been partly performed on a supercomputing system in the Supercomputing Center of Wuhan University.
Computational facilities of Radboud University (TCM/IMM) funded by the FLAG-ERA JTC2017 Project GRANSPORT are also gratefully acknowledged.
The work is partially supported by the Russian Science Foundation, Grant No. 17-72-20041.
\end{acknowledgments}

{
\appendix*
\label{apndx}
\section{Derivation of anharmonic scattering probability}

Here we provide details on the derivation of Eq.~(\ref{Vq2}). Treating the phonon fields classically, the thermal average of $\left|V_{{\bf q}}\right|^{2}$ is given by the following functional integral over in-plane and out-of-plane displacement fields:
\begin{equation}
 \langle |V_{\bf q}|^2 \rangle 
  = Z^{-1} \int Dh\left({\bf r}\right)\int D{\bf u}\left({\bf r}\right)\,  \left|V_{\bf q}\right|^{2} e^{-\beta H},
 \label{functional_integral}
\end{equation}
where the Boltzmann weight is determined by the Hamiltonian in Eq.~(\ref{hphon}). The partition function $Z$ is given by:
\begin{equation}
 Z = \int Dh\left({\bf r}\right) \int D{\bf u}\left({\bf r}\right)\,e^{-\beta H}.
 \label{partition_function}
\end{equation}
Before performing the Gaussian integration over in-plane fields, it is useful to decompose $f_{\alpha \beta}\left({\bf q}\right)$, the Fourier transform of $f_{\alpha \beta}\left({\bf r}\right) = \partial_{\alpha} h\left({\bf r}\right) \partial_{\beta}h\left({\bf r}\right)$, into longitudinal and transverse parts~\cite{Nelson, Nelson-orig}:
\begin{equation}
 \begin{split}
 f_{\alpha \beta}\left({\bf q}\right) =  i \left[q_{\alpha} \phi_{\beta}\left({\bf q}\right) + q_{\beta} \phi_{\alpha}\left({\bf q}\right)\right] + P^{T}_{\alpha \beta}\left({\bf q}\right) f^{T}\left({\bf q}\right).
\end{split}
\label{long_transverse}
\end{equation}
Here $P^{T}_{\alpha \beta}\left({\bf q}\right) = \delta_{\alpha \beta} - q_{\alpha} q_{\beta}/q^{2}$ is the transverse projector and $f^{T}\left({\bf q}\right) = P^{T}_{\gamma \delta}\left({\bf q}\right)f_{\gamma \delta}\left({\bf q}\right)$. It is then convenient to change variables in the functional integral~\cite{Nelson, Nelson-orig} and to integrate over shifted in-plane fields, defined in Fourier space as:
\begin{equation}
\tilde{u}_{\alpha}\left({\bf q}\right) \equiv u_{\alpha}\left({\bf q}\right) + \phi_{\alpha}\left({\bf q}\right) - \nu\frac{i}{2} \frac{q_{\alpha}}{q^{2}} f^{T}\left({\bf q}\right),
\label{change_variables}
\end{equation}
where $\nu = \lambda/(\lambda + 2G)$ is the 2D Poisson ratio. In terms of the variables $\tilde{u}_{\alpha}\left({\bf q}\right)$, the Hamiltonian reads
\begin{equation}
\begin{split}
 & H  = \sum_{{\bf q}} \bigg\{ \frac{1}{2} \kappa q^{4}\left|h_{{\bf q}}\right|^{2} + \frac{Y}{8} f^{T}\left({\bf q}\right) f^{T}\left(-{\bf q}\right) \\ &  + \frac{1}{2} \left[(\lambda + G) q_{\alpha} q_{\beta} + G \, q^{2} \delta_{\alpha \beta}\right] \tilde{u}_{\alpha}\left({\bf q}\right) \tilde{u}_{\beta}\left(-{\bf q}\right)\bigg\}, \\
\end{split}
\label{hphon_shifted}
\end{equation}
where $Y = \lambda \left(1- \nu^{2}\right)/\nu$. The Fourier transform of the deformation potential can be written as
\begin{equation}
 V_{{\bf q}} = i g q_{\alpha} \tilde{u}_{\alpha}\left({\bf q}\right) + \frac{g}{2} \left(1-\nu\right) f^{T}\left({\bf q}\right).
\end{equation}
The out-of-plane fields and the in-plane modes $\tilde{u}_{\alpha}\left({\bf q}\right)$ are now decoupled in the Hamiltonian. The thermal average of $\left|V_{{\bf q}}\right|^{2}$ is therefore:
\begin{equation}
 \begin{split}
\langle \left|V_{{\bf q}}\right|^{2}\rangle & = g^{2} q_{\alpha} q_{\beta} \langle \tilde{u}_{\alpha}\left({\bf q}\right) \tilde{u}_{\beta}\left(-{\bf q}\right)\rangle \\ &+ g^{2}\frac{(1 -\nu)^{2}}{4} \langle f^{T}\left({\bf q}\right) f^{T}\left(-{\bf q}\right)\rangle.
 \end{split}
 \label{average_Vq2}
\end{equation}

Since the Hamiltonian is quadratic in the fields $\tilde{u}_{\alpha}$, the correlation function $\langle \tilde{u}_{\alpha}({\bf q}) \tilde{u}_{\beta}(-{\bf q})\rangle$ can be calculated explicitly. From the Hamiltonian~(\ref{hphon_shifted}) we find:
\begin{equation}
 \langle \tilde{u}_{\alpha}\left({\bf q}\right) \tilde{u}_{\beta}\left(-{\bf q}\right)\rangle = \frac{T}{\left(\lambda + 2G\right)q^{2}} P^{L}_{\alpha \beta}\left({\bf q}\right)+ \frac{T}{G q^{2}} P^{T}_{\alpha \beta}\left({\bf q}\right),
 \label{harmonic_average}
\end{equation}
where $P^{L}_{\alpha \beta}\left({\bf q}\right) = q_{\alpha} q_{\beta}/q^{2}$ is the longitudinal projector. Combining Eqs.~(\ref{average_Vq2}) and~(\ref{harmonic_average}) yields:
\begin{equation}
\begin{split}
& \langle \left|V_{{\bf q}}\right|^{2}\rangle  = g^{2} \frac{T}{\lambda + 2 G} + g^{2} \frac{(1 -\nu)^{2}}{4} \langle f^{T}\left({\bf q}\right) f^{T}\left(-{\bf q}\right)\rangle\\
 & = g^{2} \frac{T}{Y}\left(1-\nu^{2}\right) + g^{2} \frac{(1 -\nu)^{2}}{4} \langle f^{T}\left({\bf q}\right) f^{T}\left(-{\bf q}\right)\rangle,
\end{split}
 \end{equation}
which is Eq.~(\ref{Vq2}) in the main text.
}

\end{document}